\documentclass[10pt,conference]{IEEEtran}
\usepackage{cite}
\usepackage{amsmath,amssymb,amsfonts}
\usepackage{algorithmic}
\usepackage{graphicx}
\usepackage{textcomp}
\usepackage{xcolor}
\usepackage[hyphens]{url}
\usepackage{fancyhdr}
\usepackage{hyperref}
\usepackage{booktabs}

\title{ComFuse: Fusing Complex Memory-Intensive Subgraphs with Compute-Intensive Kernels For Modern GPU Architectures}

\newcommand\hpcaauthors{Di Mu$\dagger$, Tengyuan Jin$\dagger$, Zhenkun Wang$\dagger$, Jialin Yang $\ddagger$, YuSen Li$\dagger$, \\Mian Huo$\ddagger$, Shusong Guo$\ddagger$, Gang Wang$\dagger$, Xiaoguang Liu$\dagger$}
\newcommand\hpcaaffiliation{Nankai University $\dagger$, ByteDance$\ddagger$}
\newcommand\hpcaemail{\{mudi, jinty, liyusen, wgzwp, liuxg\}@nbjl.nankai.edu.cn\\\{yangjialin, huomian, guoshusong\}@bytedance.com\\kun51436@gmail.com}

\author{
    \IEEEauthorblockN{\hpcaauthors{}}
      \IEEEauthorblockA{
        \hpcaaffiliation{} \\
        \hpcaemail{}
      }
}

\begin{document}
\maketitle

\thispagestyle{plain}
\pagestyle{plain}

\newcommand{\hpcaheight}{0mm}


\begin{abstract}
Modern deep learning workloads increasingly comprise heterogeneous computation graphs that combine compute-intensive operators with memory-intensive subgraphs. Existing deep learning compilers typically optimize these operator classes separately, creating rigid fusion boundaries that limit cross-operator optimization and on-chip data reuse. We observe that downstream memory-intensive operations can execute concurrently with compute-intensive operators, allowing their execution to be hidden behind computation; however, automatically exploiting this opportunity poses new compilation challenges.

In this paper, we present \textit{ComFuse}, an automated GPU compilation system that employs a novel operator fusion strategy to generate high-performance kernels for complex graph structures comprising compute-intensive operators and dependency-rich, memory-intensive elementwise–reduction subgraphs. \textit{ComFuse} further supports the fusion of back-to-back GEMM (B2BGEMM) patterns, extending its applicability to more complex compute–memory interaction patterns. Additionally, it automatically lowers high-level tensor subprograms into optimized fused kernels, reducing the need for manual kernel engineering. Experimental results show that the fused kernels generated by \textit{ComFuse} outperform those produced by TorchInductor across post-norm workloads and various complex computation scenarios, while supporting more flexible fusion patterns.
\end{abstract}

\section{Introduction}
\label{sec:introduction}
The rapid growth of deep learning workloads, such as recommendation models and large language models (LLMs), has led to increasingly complex computation graphs composed of both compute-intensive and memory-intensive operators\cite{dlrm2019, wolters2024memory, li2024onednn}.
Among these operators, tensor contractions such as matrix multiplication (MatMul) often account for a substantial fraction of floating-point computation, whereas ElementWise and Reduction operators perform tensor transformations and statistical processing.
Efficient execution of these heterogeneous operators has become a fundamental challenge in modern deep learning training and inference systems\cite{zheng2023chimera}.

Existing deep learning frameworks and compilers~\cite{ansel2024pytorch, abadi2016tensorflow, nvidia_tensorrt}  adopt different optimization strategies for operators with distinct performance characteristics. 
For compute-intensive operators such as MatMul, compilers exploit specialized matrix units, e.g., NVIDIA Tensor Cores and Google TPU Matrix Units, through tiling, thread mapping, and layout transformations to improve data reuse and hardware utilization\cite{yan2020demystifying, jouppi2017datacenter, gupta2024mixedinput, zheng2023chimera}. 
Meanwhile, memory-intensive operators such as ElementWise and Reduction are often bottlenecked by kernel launch overheads and data movement\cite{ivanov2021data, he2022brrrrfromfirstprinciples}. 
To address these overheads, compilers commonly apply operator fusion, which merges consecutive operators into a single kernel to reduce intermediate tensor materialization and memory traffic\cite{electronics15051034, qiao2018automatic}.

Although existing optimization strategies have achieved substantial performance improvements within their respective domains, they typically maintain an explicit materialization boundary between compute-intensive and memory-intensive operators. 
As illustrated in Figure~\ref{fig:ComFuse-overview}, intermediate results produced by MatMul are commonly materialized to global memory (GMEM) and later reloaded by downstream ElementWise or Reduction kernels for further processing. 
This on-chip--off-chip--on-chip data movement prevents downstream operators from exploiting the short-lived temporal locality of intermediate values, leading to substantial memory bandwidth overhead. 
Eliminating this boundary offers two-fold advantages. 
First, it allows downstream operators to consume MatMul outputs while they remain resident on-chip, thereby significantly reducing GMEM traffic and maximizing on-chip data reuse. 
Second, it exploits the complementary hardware resource demands of the two classes of operators: compute-intensive kernels primarily rely on specialized matrix units (e.g., Tensor Cores), whereas memory-intensive subgraphs mainly utilize general-purpose CUDA cores for lightweight computations. 
This complementarity creates unique opportunities to overlap downstream memory-intensive processing with tensor-core computation.
We refer to the problem of jointly fusing compute-intensive operators with memory-intensive subgraphs as \textbf{\textit{joint compute–memory operator fusion}}.
However, realizing such fusion is nontrivial because the two operator classes differ substantially in execution granularity, data organization, and synchronization requirements. Reduction operations further introduce dependency patterns that may span multiple elements, tiles, or execution units, preventing straightforward producer–consumer fusion.

To this end, we propose \textit{ComFuse}, a compilation system that automatically generates high-performance fused kernels for subgraphs combining compute-intensive kernels and memory-intensive operations. 
\textit{ComFuse} leverages modern GPU architecture support for on-chip cooperation and introduces a Stage-Stream Execution Model that orchestrates fine-grained data movement across computation stages, enables complex memory-intensive subgraphs, such as deeply nested Reductions interleaved with ElementWise operations, to be fused with compute-intensive operators.
To support more complex multi-MatMul structures, \textit{ComFuse} further introduces a B2BGEMM scheduling paradigm that fuses two MatMul operations connected by an intervening memory-intensive subgraph.
The main contributions of this work are as follows:

\begin{figure}[t]
  \centering
  \includegraphics[width=\columnwidth]{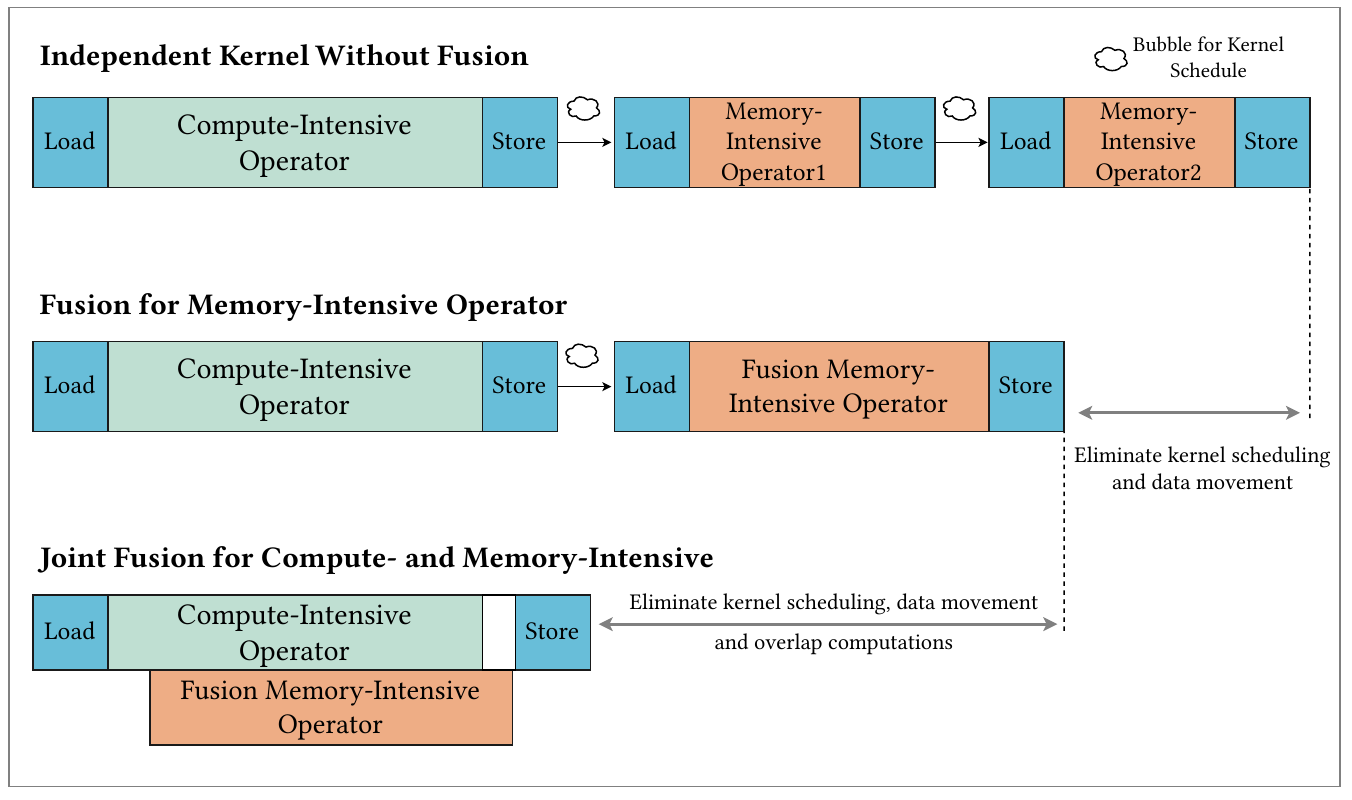}
  \caption{Joint Fusion of Compute-intensive and Memory-intensive Operators Across Data Boundaries.}
  \label{fig:ComFuse-overview}
\end{figure}

\begin{enumerate}
  \item We identify the boundary between compute-intensive and memory-intensive operators as a key barrier to on-chip data reuse, and propose joint compute--memory fusion to eliminate unnecessary data movement.
  \item We introduce a Stage-Stream Execution Model that enables fine-grained fusion of compute-intensive kernels with complex memory-intensive subgraphs involving ElementWise and Reduction operations.
  \item We develop a B2BGemm scheduling paradigm to support joint fusion across more complex multi-MatMul structures.
  \item We implement \textit{ComFuse}, a compilation system that automatically captures the computational semantics of tensor subprograms and constructs them into high-performance fused kernels.
\end{enumerate}

\section{Related Work and Background}
\label{sec:background-related-work}
\subsection{Toward Joint Compute-Memory Fusion}
Modern deep learning systems have extensively optimized compute-intensive kernels through automatic scheduling, hardware-aware programming abstractions, or vendor-tuned libraries~\cite{adams2019learning, sivathanu2019astra, chen2018tvm, chen2018learning, zheng2020ansor, shao2022metaschedule, triton, hagedorn2020fireiron, ansel2024pytorch, nvidia_cublas, chetlur2014cudnn, zhang2024mcfuser}. 
However, by primarily optimizing the core compute-intensive kernels and their local schedules, these approaches leave fusion opportunities with complex memory-intensive subgraphs largely unexplored.

For memory-intensive operators, fusion is the dominant strategy for reducing kernel launch overhead and intermediate memory traffic. 
Early compiler systems such as XLA~\cite{tensorflow_xla}, TensorRT~\cite{nvidia_tensorrt}, Tensor Comprehensions~\cite{vasilache2018tensor}, and TVM-based frameworks~\cite{chen2018tvm, shao2022metaschedule, zheng2023operator} mainly target common memory-intensive patterns, such as ElementWise chains and limited Reduction operations, using heuristic rules, templates, or pattern matching. 
Later works, such as DNNFusion~\cite{niu2021dnnfusion}, Apollo~\cite{zhao2022apollo}, FusingStitch~\cite{zheng2020fusionstitching}, and AStitch~\cite{zheng2022astitch}, further target more complex memory-intensive fusion.

Recent systems have begun to explore joint compute--memory fusion from several directions.
CUTLASS~\cite{NVIDIA2017cutlass} with EVT~\cite{chen2024evt} and Bolt~\cite{xing2022bolt} investigate fusion between compute-intensive kernels and simple memory-intensive operations. 
In a complementary direction, PluS~\cite{wu2025plus} extends template matching with loop-structure-based matching, broadening the applicability of handwritten fused kernels. 
Mirage~\cite{wu2025mirage} and FlashLight~\cite{you2025flashlight} rely on algebraic transformations to optimize structured computation patterns. 
CODA~\cite{guo2026coda} applies algebraic transformations to Transformer computations, further highlighting the importance of joint compute--memory fusion.

However, these approaches remain limited when the memory-intensive region contains dependency-rich reductions, multi-level aggregation, or nontrivial producer--consumer relationships with the compute-intensive kernel.
Algebraic transformations and template matching impose strong structural constraints on fusible subgraphs, while manually written fused kernels are difficult to adapt to rapidly evolving model architectures. 
Addressing this gap motivates \textit{ComFuse}, which targets automatic joint fusion of compute-intensive kernels with dependency-rich memory-intensive subgraphs.

\subsection{Technical Challenges for Joint Fusion}
\label{Technical Challenges}
\subsubsection{Challenge 1: Fusing Memory-Intensive Operators with Diverse Parallel Semantics}
Elementwise operations exhibit regular element-level parallelism: each output element depends only on the corresponding elements of its input tensors.
As a result, chains of ElementWise operations can often be fused in a straightforward producer-consumer manner without materializing intermediate results in GMEM. 
In contrast, reduction operations rely on the aggregation of multiple elements and typically require data exchange and synchronization across compute units. 
Reduction outputs are often broadcast to subsequent ElementWise operations, forming a dependency pattern of multi-element aggregation followed by broadcast. 
This pattern breaks the simple element-to-element dataflow and makes it difficult to fuse dependency-rich ElementWise-Reduction subgraphs without carefully coordinating data movement and synchronization.

\subsubsection{Challenge 2: Aligning Tiled Execution with Reduction Dependencies}
\begin{figure}[t]
  \centering
  \includegraphics[width=\columnwidth]{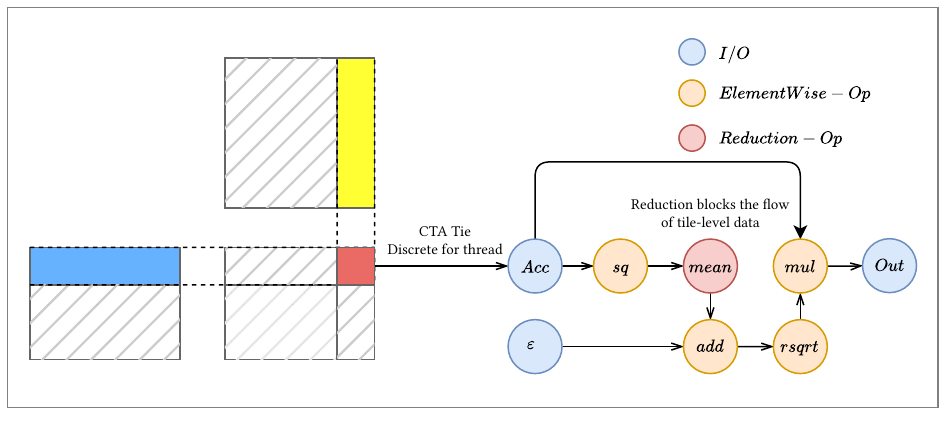}
  \caption{Data Flow in MatMul + RMSNorm Subgraph Structure}
  \label{fig:matmul-rmsnorm-dataflow}
\end{figure}

Compute-intensive operators such as MatMul typically employ multi-level tiling, with each cooperative thread array (CTA) incrementally producing output tiles. Elementwise operations can immediately consume these tiles on-chip because they preserve the tile-level coordinate mapping. Reductions, however, may aggregate values across multiple tiles, creating a mismatch between tiled production and cross-tile dependencies.
As illustrated in Figure~\ref{fig:matmul-rmsnorm-dataflow}, in a MatMul + RMSNorm subgraph, \texttt{square} can process each output tile as soon as it is produced, whereas \texttt{mean} must wait for multiple tiles and therefore blocks the streaming dataflow. Since downstream normalization still requires the original MatMul outputs, limited on-chip capacity may force them to be materialized in GMEM and reloaded after the reduction. Supporting cross-tile reductions without disrupting tiled execution or introducing unnecessary materialization is therefore a key challenge for joint compute--memory fusion.

\subsubsection{Challenge 3: Automating Fusion for Diverse Operator Compositions}

Modern deep learning models contain diverse compositions of elementwise operations, reductions, normalizations, activations, and scaling operations, giving rise to a large and rapidly evolving space of subgraph patterns. These patterns vary in topology, dependency structure, tensor shape, and operator ordering as model architectures evolve.

Fixed-pattern fusion and handwritten kernels require substantial development and maintenance effort, as new subgraph variants may require revisiting execution schedules, data layouts, synchronization strategies, and memory management. Moreover, such approaches generalize poorly to previously unseen compositions. Addressing this diversity therefore requires an automated compiler framework that can analyze computational semantics and generate high-performance fused kernels for a broad range of subgraph structures.

\begin{figure*}[t]
  \centering
  \includegraphics[width=\textwidth]{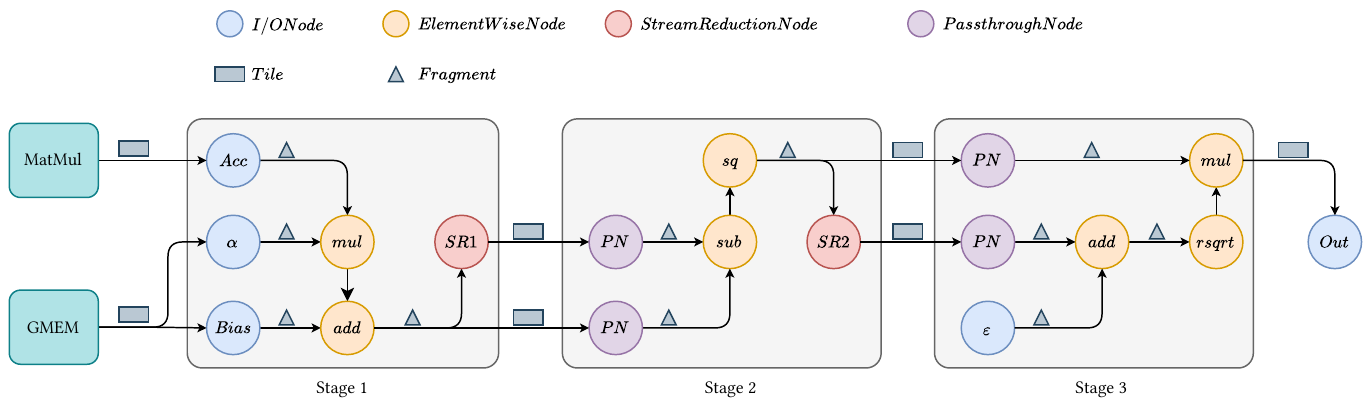}
  \caption{Case for The Stage-Stream Execution Model: MatMul + LayerNorm}
  \label{fig:3_1}
\end{figure*}

\subsection{Architectural Opportunities}
\label{sec:cluster-constraints}

Recent NVIDIA GPU architectures, including Hopper and Blackwell, introduce Thread Block Clusters to support cooperative execution across CTAs~\cite{cuda_programming_guide}. In the conventional CTA-isolated execution model, efficient cross-CTA communication and synchronization are limited, often requiring intermediate values crossing CTA or tile boundaries to be materialized in GMEM.

Thread Block Clusters allow multiple CTAs to be co-scheduled within the same GPU Processing Cluster (GPC). Their Distributed Shared Memory (DSMEM) exposes the shared memory of clustered CTAs as a distributed address space, while cluster-level synchronization enables coordinated cross-CTA execution. Together, these mechanisms provide temporal coordination and spatial data sharing, allowing intermediate values to be exchanged on-chip across CTA boundaries.

This architectural support creates new opportunities for joint compute--memory fusion, particularly when downstream elementwise and reduction operations consume MatMul outputs spanning multiple tiles. \textit{ComFuse} incorporates these cross-CTA capabilities into operator fusion through its Stage-Stream Execution Model and B2BGEMM scheduling paradigm, enabling more general and efficient fused execution on modern GPUs.

\section{The Stage-Stream Execution Model}
\label{sec:Stage-Stream Execution Model}

As discussed in Sec~\ref{Technical Challenges}, Reduction operations inherently require data synchronization. 
In conventional execution models, achieving this synchronization typically forces compilers to explicitly materialize the entire intermediate tensor in GMEM. 
However, because most epilogue reductions aggregate data along only one logical dimension, the actual synchronization boundary is confined to a local subset of tiles rather than the entire tensor.
Specifically, under a two-dimensional tiling scheme, the intermediate tensor $P$ is partitioned into a set of tiles:
\begin{equation}
  P = \{T_{m,n} \mid 0 \le m < M_T,\ 0 \le n < N_T\},
\end{equation}
where $m$ and $n$ denote the tile indices along the non-reduction and reduction dimensions, respectively. 
Since the reduction depends solely on the tile subset $\{T_{m,n}\}_{n=0}^{N_T-1}$ sharing the same non-reduction index $m$, the synchronization boundary of a single reduction is strictly confined to this local tile subset, eliminating the need for global tensor materialization.

Based on this observation, we design the \textbf{Stage-Stream Execution Model}, as illustrated in Figure\ref{fig:3_1}. 
This model uses Reduction operations as stage boundaries and decomposes a complete memory-intensive subgraph into multiple sequentially executed Directed Acyclic Graphs(DAG). Each DAG corresponds to one execution stage: within a stage, local ElementWise and Reduction are performed; across stages, DAGs are connected through reduction results and bypassed intermediate data. The remainder of this section introduces the intra-DAG computation, inter-DAG chaining, and the pipeline scheduling mechanism that coordinates the Reduction execution with the MatMul.

\subsection{Intra-DAG computation}
We represent the $i$-th stage as a DAG subgraph $G_i = (V_i, E_i)$, where $V_i$ denotes the set of computation nodes (including ElementWise and Reduction nodes) and $E_i$ represents the logical data dependencies. 
Unlike conventional compiler IRs, where an edge typically implies a buffer materialized in GMEM, an edge $d \in E_i$ represents only a logical data dependency that can be mapped to a register-file data channel.

The execution of a stage DAG consists of two sequential phases: fine-grained \texttt{compute} and tile-level \texttt{reduce}. 
During the \texttt{compute} phase, data propagates along the dependency edges $d \in E_i$ at the granularity of a \textit{Fragment} (the largest group of elements processed by a single vectorized instruction). 
While Fragments can stream seamlessly through element-wise nodes, directly passing them through reduction nodes yields incorrect semantics due to cross-thread dependencies. 
Instead, reduction nodes intercept the stream to perform thread-local aggregation over the Fragment elements, confining the reduction to the thread level before global aggregation.

\begin{figure}[t]
  \centering
  \includegraphics[width=\columnwidth]{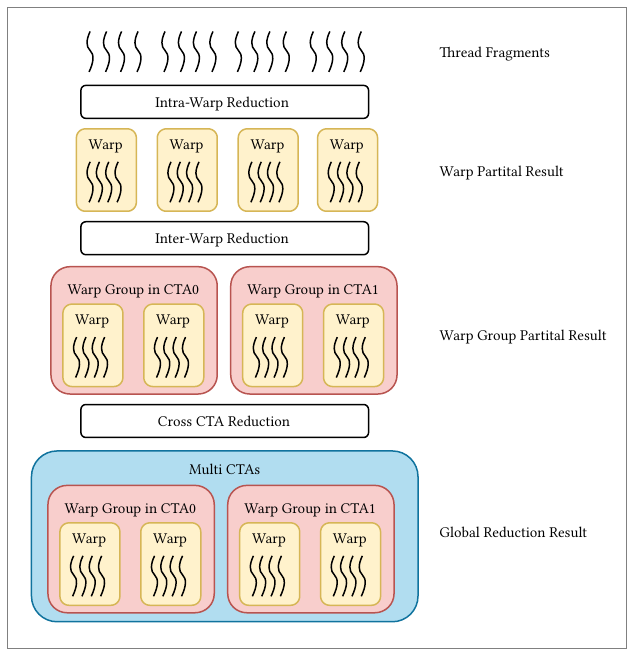}
  \caption{Three-level Reduction Scheme in Stage-Stream Execution Model.}
  \label{fig:3_2}
\end{figure}

During the \texttt{reduce} phase, the reduction node performs hierarchical aggregation to resolve cross-thread dependencies. 
Building upon the thread-local results, \textit{ComFuse} employs a three-level reduction scheme(Figure~\ref{fig:3_2}): (1) \textit{intra-warp reduction} via register-level butterfly shuffles, (2) \textit{intra-warp-group reduction} via SMEM, and (3) \textit{inter-CTA aggregation} along the logical reduction dimension.
To optimize the third level, \textit{ComFuse} introduces a dual-mode adaptive aggregation strategy based on the number of participating CTAs ($C_N$). 
When $C_N$ is small, \textit{ComFuse} employs a peer-to-peer cross-broadcast strategy, where each CTA performs decentralized all-to-all broadcasting over DSMEM to independently complete the final aggregation locally, thereby avoiding the synchronization bottlenecks of centralized coordination. 
Conversely, when $C_N$ is large, \textit{ComFuse} switches to a leader-follower aggregation strategy to prevent DSMEM congestion. 
In this mode, one CTA is elected as the leader to perform centralized aggregation; follower CTAs send their local results to the leader, which computes the global reduction and broadcasts the final result back. 
This adaptive hierarchical scheme minimizes inter-CTA communication latency and instruction overhead while ensuring global data consistency.

To support the aforementioned inter-CTA aggregation, \textit{ComFuse} leverages the spatial and temporal constraints of Thread Block Clusters introduced in Sec~\ref{sec:cluster-constraints}. 
Specifically, \textit{ComFuse} uses DSMEM as the on-chip communication path for inter-CTA data exchange, mapping all CTAs participating in a single reduction operation to the same Cluster. 
By scheduling tasks at the Cluster granularity and applying a secondary task mapping based on each CTA's relative offset within the Cluster, this mechanism guarantees that cooperating CTAs are co-scheduled concurrently in both space and time. 
Consequently, any CTA can retrieve and aggregate partner results via DSMEM while its own intermediate values still reside in registers, enabling the dataflow within a CTA to stream through the reduction node at tile granularity with correct global reduction results.

This highly efficient inter-CTA aggregation is enabled by two key designs. 
The first is \textit{cross-CTA data layout consistency}. 
Since the mapping between logical matrix coordinates and the thread-value layout is uniform across warp groups, threads with identical IDs in different CTAs always hold the same logical rows. 
Consequently, communication endpoints can directly interpret DSMEM data using a unified indexing rule, completely eliminating the overhead of data reordering, encoding, or decoding. 
The second is \textit{TMA-enabled bulk data movement}. 
Instead of relying on conventional thread-level load/store instructions for fine-grained copies, \textit{ComFuse} leverages Tensor Memory Accelerator (TMA) instructions to perform direct bulk transfers between source and destination DSMEM addresses. 
This mechanism dramatically reduces instruction overhead and maximizes the bandwidth utilization of the DSMEM interconnect, further accelerating inter-CTA data exchange.

\subsection{Inter-DAG chaining}
Within a single stage DAG, the dataflow propagates at Fragment granularity through ElementWise nodes during the \texttt{compute} phase, and completes aggregation at tile granularity within Reduction nodes during the \texttt{reduce} phase. 
To chain multiple stage DAGs together, a Reduction node not only serves as the terminal node of the current stage, but also manages the lifecycle of the subsequent DAG, referred to as the PostOp, and drives its \texttt{compute} and \texttt{reduce} phases.

Specifically, in the DAG of Stage $i$, after the \texttt{reduce} phase, the terminal Reduction node broadcasts the aggregated reduction result to match the tile dimensions. 
This broadcasted tile is then decomposed again into Fragments and forwarded as input to the DAG of Stage $i+1$, thereby triggering the execution of the subsequent stage. 
In this way, the Reduction node establishes an explicit dataflow control boundary between adjacent stages, enabling multiple DAGs to execute in a stage-ordered, pipelined manner.

While chaining multiple stages enables complex computations, it also introduces resource challenges, particularly regarding SMEM consumption during inter-CTA reduction. 
To mitigate this, we leverage the sequential execution semantics of multi-stage DAGs. 
Since at most one Reduction node actively utilizes the SMEM buffer at any given moment, \textit{ComFuse} reuses a single, unified SMEM buffer across all Stage DAGs. 
This optimization reduces the buffer space complexity from $O(L)$ to $O(1)$, where $L$ denotes the total number of reduction nodes in the epilogue computation. 
Consequently, the Stage-Stream Execution Model can support complex epilogue computations with an arbitrary number of reduction operations without incurring linear growth in on-chip storage overhead.

Beyond buffer optimization, adjacent stage DAGs often present opportunities for reusing local intermediate results. 
For example, as shown in Figure\ref{fig:3_1}, Stage 1 computes $x = \alpha \cdot Acc + Bias$. 
This intermediate value is consumed within the current stage to compute the $mean$ reduction, but is also required by Stage 2 to evaluate $x - \text{mean}$. 
Recomputing $x$ in the subsequent stage would incur redundant computational overhead. 
To eliminate this redundancy, \textit{ComFuse} establishes a bypass data path between adjacent stages to propagate reusable intermediate values. 
Specifically, each stage DAG reserves placeholder nodes to ingest bypassed data from its predecessor. 
When the terminal Reduction node triggers the execution of the PostOp, these bypassed values are injected into the subsequent DAG alongside the reduction results at Fragment granularity.

\subsection{Pipeline Schedule}
From a macro-execution perspective, the Stage-Stream Execution Model for memory-intensive subgraphs exhibits tile-level streaming semantics. 
Following the terminology established in CUTLASS~\cite{NVIDIA2017cutlass}, we define the \textit{MainLoop} as the core computation of the compute-intensive matrix multiplication, and the \textit{Epilogue} as the general computation of the memory-intensive subgraphs. 
In each iteration, the MainLoop produces an Accumulator Tile, which is subsequently consumed by the Epilogue on a tile-by-tile basis. 
Since the \textit{MainLoop} and the \textit{Epilogue} rely on different hardware resources, their executions can be overlapped, as PingPong Gemm.
\textit{ComFuse} adopts a PingPong Pipeline scheduling strategy: while a CTA is executing the \textit{MainLoop} computation for one tile, it simultaneously performs \textit{Epilogue} processing for another tile. 
By interleaving \textit{MainLoop} and \textit{Epilogue} execution across tiles, \textit{ComFuse} can hide the \textit{Epilogue} latency behind the \textit{MainLoop} under ideal workload conditions.

\section{Fusion for B2BGEMM}
\label{sec:Fusion for B2BGEMM}
The Stage-Stream Execution Model addresses the fusion of a compute-intensive operator with a memory-intensive subgraph. Building upon this model, \textit{ComFuse} further explores the fusion of multiple compute-intensive tasks, namely the B2BGEMM. Its general computation can be formulated as
\begin{equation}
  P = XW_1,\quad S = F(P),\quad O = SW_2,
\end{equation}
Here, $X \in \mathbb{R}^{M \times K}$ denotes the input matrix, while $W_1 \in \mathbb{R}^{K \times N_1}$ and $W_2 \in \mathbb{R}^{N_1 \times N_2}$ denote the weight matrices of the two MatMul operations, respectively. $P \in \mathbb{R}^{M \times N_1}$ denotes the raw intermediate result produced by the first MatMul, $F(\cdot)$ represents the intermediate memory-intensive operation between the two MatMuls, $S = F(P)$ denotes the post-processed intermediate matrix, and $O \in \mathbb{R}^{M \times N_2}$ denotes the final output matrix.

B2BGEMM is a ubiquitous computation pattern in deep learning. 
Conventional implementations typically decompose B2BGEMM into separate kernels (i.e., $\mathrm{MatMul}_1$, an intermediate $\mathrm{Epilogue}$, and $\mathrm{MatMul}_2$), which forces the materialization of intermediate results in GMEM and incurs severe memory traffic overhead. 

The primary barrier to fusing B2BGEMM lies in the granular mismatch between the tile production of $\mathrm{GEMM}_1$ and the tile consumption of $\mathrm{GEMM}_2$. 
Specifically, $\mathrm{GEMM}_1$ partitions and computes the intermediate matrix in parallel across different CTAs, leaving each CTA with only a local tile of the intermediate result. 
In contrast, producing an output tile in $\mathrm{GEMM}_2$ requires a dot-product accumulation that spans multiple contiguous intermediate tiles along the contraction dimension $N_1$. 
This dependency forces conventional systems into a performance dilemma: they must either redundantly recompute intermediate tiles to preserve data locality, or fully materialize them in GMEM to allow $\mathrm{GEMM}_2$ to reload the required data. 

\begin{figure}[t]
  \centering
  \includegraphics[width=\columnwidth]{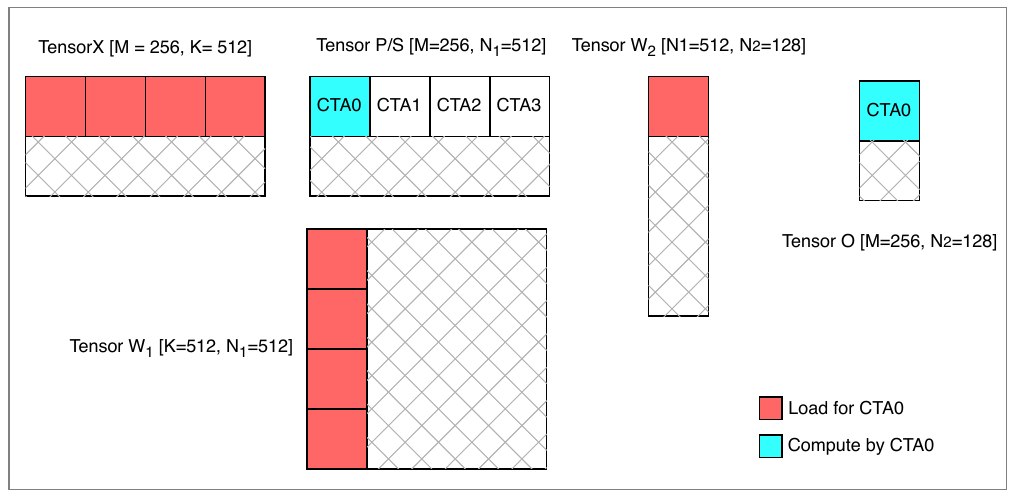}
  \caption{B2BGEMM Parallel Task Schedule. }
  \label{fig:4_1}
\end{figure}
\subsection{Dataflow Scheduling}
To reduce the memory access overhead caused by intermediate results in B2BGEMM, \textit{ComFuse} introduces a cluster-cooperative dataflow scheduling mechanism. This mechanism allows the intermediate tiles of $S$ to be directly consumed through registers and on-chip communication paths, without being materialized in GMEM.
As illustrated in Figure~\ref{fig:4_1}, 
during the $\mathrm{MatMul}_1$ stage, \textit{ComFuse} partitions the workload along the output space of the intermediate matrix $P \in \mathbb{R}^{M \times N_1}$. Each CTA is responsible for computing one local tile, denoted as $P_{\text{tile}}$. The intermediate epilogue operation $F(\cdot)$ is then fused into $\mathrm{MatMul}_1$ through the Stage-Stream Execution Model. After the Epilogue computation, each CTA transforms its local $P_{\text{tile}}$ into the corresponding post-processed tile:
\begin{equation}
  S_{\text{tile}} = F(P_{\text{tile}}).
\end{equation}
At this point, the intermediate tile $S_{\text{tile}}$ exhibits two key properties that enable on-chip consumption. 
Microscopically, $S_{\text{tile}}$ remains strictly resident in thread registers, completely bypassing the latency and bandwidth overhead of SMEM or GMEM write-backs. 
Macroscopically, CTAs aligned along the contraction dimension $N_1$ within the same Thread Block Cluster hold logically contiguous segments of the intermediate matrix $S$. 
This spatial alignment, combined with the spatiotemporal co-scheduling of the Cluster, allows these cooperating CTAs to seamlessly transition into the subsequent $\mathrm{MatMul}_2$ stage without losing data locality.
During the $\mathrm{MatMul}_2$ stage, \textit{ComFuse} exploits this on-chip layout by cooperatively distributing the computation of a single output tile $O_{\text{tile}}$ across the CTAs within the Cluster. 
Specifically, each participating CTA leverages its register-resident $S_{\text{tile}}^{(c)}$ alongside the corresponding weight tile $W_{2,\text{tile}}^{(c)}$ (preloaded into SMEM) to compute a partial accumulation:
\begin{equation}
  O_{\text{tile}}^{(c)} = S_{\text{tile}}^{(c)} W_{2,\text{tile}}^{(c)},
\end{equation}
where $c$ denotes the local CTA index within the Cluster. 
Because these partial results represent segmented accumulations along the contraction dimension $N_1$, they are aggregated during the final global memory commit to produce the complete output tile:
\begin{equation}
  O_{\text{tile}} = \sum_{c \in \mathcal{C}} O_{\text{tile}}^{(c)},
\end{equation}
where $\mathcal{C}$ represents the set of co-scheduled CTAs cooperating on the same output tile.

By maintaining $S_{\text{tile}}$ entirely within registers and on-chip data paths throughout the B2BGEMM pipeline, \textit{ComFuse} directly feeds the outputs of $\mathrm{MatMul}_1$ into $\mathrm{MatMul}_2$. 
Compared with conventional kernel-level decomposition, this cooperative scheduling paradigm completely eliminates the GMEM materialization and reload of the intermediate matrix, thereby drastically reducing memory bandwidth pressure and maximizing hardware efficiency.

\subsection{Producer Warp Group}
Within each CTA, the Producer Warp Group is responsible for prefetching the data required by $\mathrm{MatMul}_1$, the intermediate Epilogue, and $\mathrm{MatMul}_2$ into a multi-stage circular buffer in SMEM at the tile granularity. In B2BGEMM, allocating separate SMEM buffers for $\mathrm{MatMul}_1$ and $\mathrm{MatMul}_2$ would significantly increase on-chip storage consumption. To avoid this overhead, \textit{ComFuse} reuses the same set of multi-stage SMEM buffers across the two MatMul operations.
Consider the matrix dimensions
\begin{equation}
  X \in \mathbb{R}^{M \times K},\quad
  W_1 \in \mathbb{R}^{K \times N_1},\quad
  W_2 \in \mathbb{R}^{N_1 \times N_2}.
\end{equation}
Let the CTA-level tile shape be $\mathrm{TileM} \times \mathrm{TileN} \times \mathrm{TileK}$.
In $\mathrm{MatMul}_1$, the computation is reduced along the $K$ dimension and therefore requires $K_1 = K / \mathrm{TileK}$ iterations. 
In each iteration, the Producer loads one $X_{\text{tile}} \in \mathbb{R}^{\mathrm{TileM} \times \mathrm{TileK}}$ and one $W_{1,\text{tile}} \in \mathbb{R}^{\mathrm{TileK} \times \mathrm{TileN}}$ into SMEM.
In $\mathrm{MatMul}_2$, the computation is reduced along the intermediate dimension $N_1$, and hence requires $K_2 = \mathrm{TileN} / \mathrm{TileK}$ iterations. 
Since the intermediate matrix tile $S_{\text{tile}}$ is directly produced by $\mathrm{MatMul}_1$ and the Epilogue in registers, each iteration of $\mathrm{MatMul}_2$ only needs to prefetch one $W_{2,\text{tile}} \in \mathbb{R}^{\mathrm{TileK} \times \mathrm{TileN}}$ from GMEM into SMEM.
We observe that $W_{1,\text{tile}}$ and $W_{2,\text{tile}}$ require the same SMEM buffer structure. Based on this observation, \textit{ComFuse} constructs a unified multi-stage circular SMEM buffer using the buffer configuration of $\mathrm{MatMul}_1$, and alternately loads the $X_{\text{tile}}$ and $W_{1,\text{tile}}$ tiles required by $\mathrm{MatMul}_1$, as well as the $W_{2,\text{tile}}$ tiles required by $\mathrm{MatMul}_2$. As a result, the complete SMEM buffer overhead of B2BGEMM remains on the same order as that of a single MatMul, rather than increasing linearly with the number of MatMul operations.

\subsection{Consumer Warp Group}
Within each CTA, the Consumer Warp Group is responsible for executing $\mathrm{MatMul}_1$, the intermediate Epilogue, $\mathrm{MatMul}_2$, and the final write-back. 
For a single MatMul, the original two-stage PingPong schedule between the MainLoop and the Epilogue is sufficient to hide the overhead of memory-intensive post-processing. However, in B2BGEMM, the execution flow consists of two compute-intensive MatMuls, with an intermediate Epilogue stage that mixes memory access and reduction operations. Therefore, the original two-stage PingPong model is no longer adequate to capture the pipeline dependencies and resource overlap opportunities in the two-layer MatMul structure.
\textit{ComFuse} restructures the Consumer execution into a four-stage pipeline:
\begin{equation}
  \mathrm{ML}_1
  \rightarrow
  \mathrm{Epi}_1
  \rightarrow
  \mathrm{ML}_2
  \rightarrow
  \mathrm{Epi}_2.
\end{equation}
Here, $\mathrm{ML}_1$ computes the local accumulation result $P_{\text{tile}}$ of the first MatMul; $\mathrm{Epi}_1$ applies the intermediate post-processing operation $F(\cdot)$ to $P_{\text{tile}}$ and generates the register-resident $S_{\text{tile}}$; $\mathrm{ML}_2$ consumes $S_{\text{tile}}$ together with $W_{2,\text{tile}}$ to compute the partial result of the output tile; and $\mathrm{Epi}_2$ completes the final aggregation and commits the result to GMEM.
Within the same CTA, different Consumer Warp Groups are staggered by one pipeline stage according to the above four-stage schedule. Specifically, when one Consumer Warp Group is executing $\mathrm{ML}_1$, another can execute $\mathrm{Epi}_1$, while others may simultaneously advance $\mathrm{ML}_2$ or $\mathrm{Epi}_2$. 

\section{ComFuse Implement}
\label{sec:ComFuse-design}
We implement the Stage-Stream Execution Model and the B2B-GEMM fusion mechanism described in Sec~\ref{sec:Stage-Stream Execution Model} and Sec~\ref{sec:Fusion for B2BGEMM} on top of CUTLASS. To express streaming reduction semantics within CUTLASS, we introduce a new \texttt{StreamReduction} node into the EVT node primitive system, formalized as
\begin{equation}
  \mathrm{SR\ Node} = \langle \texttt{PostOp}, \texttt{ReduceOp}, \texttt{ReduceMode}, K \rangle.
\end{equation}
Here, \texttt{PostOp} denotes the successor DAG Visitor Node attached to the current Reduction node; \texttt{ReduceOp} specifies the core reduction operator, including common mathematical primitives such as \texttt{sum}, \texttt{prod}, \texttt{max}, and \texttt{min}; \texttt{ReduceMode} extends the basic reduction semantics with specialized modes such as \texttt{mean} and \texttt{softmax}; and the hyperparameter $K$ parameterizes more complex reduction scenarios, such as Top-K reduction. 
For the B2BGEMM fusion paradigm, we further extend the Collective scheduling component in CUTLASS, enabling the joint fusion of multiple MatMul operations combined with memory-intensive subgraphs.

Although CUTLASS leverages the Epilogue Visitor Tree (EVT) to model post-MatMul operations as composable nodes, its programming interface relies heavily on template metaprogramming. 
This low-level implementation requires complex template instantiation and manual runtime parameter packing, creating a severe semantic gap with the high-level tensor-algebraic representations in deep learning frameworks. 
Consequently, integrating EVT into automated compilation workflows remains highly challenging.

To bridge this gap, we design and implement \textit{ComFuse}, a compilation system that automatically translates high-level tensor subprogram into high-performance CUTLASS kernel templates. 
Unlike manual template assembly, \textit{ComFuse} features a fully automated compilation pipeline that performs Epilogue construction and runtime parameter organization. 
By generating optimized kernels based on a unified tensor algebra representation and scheduling rules rather than fixed operator patterns, \textit{ComFuse} naturally generalizes to diverse subgraph variants, bypassing the brittleness of pattern-matching compilers against structural variations.

\begin{figure}[t]
  \centering
  \includegraphics[width=\columnwidth]{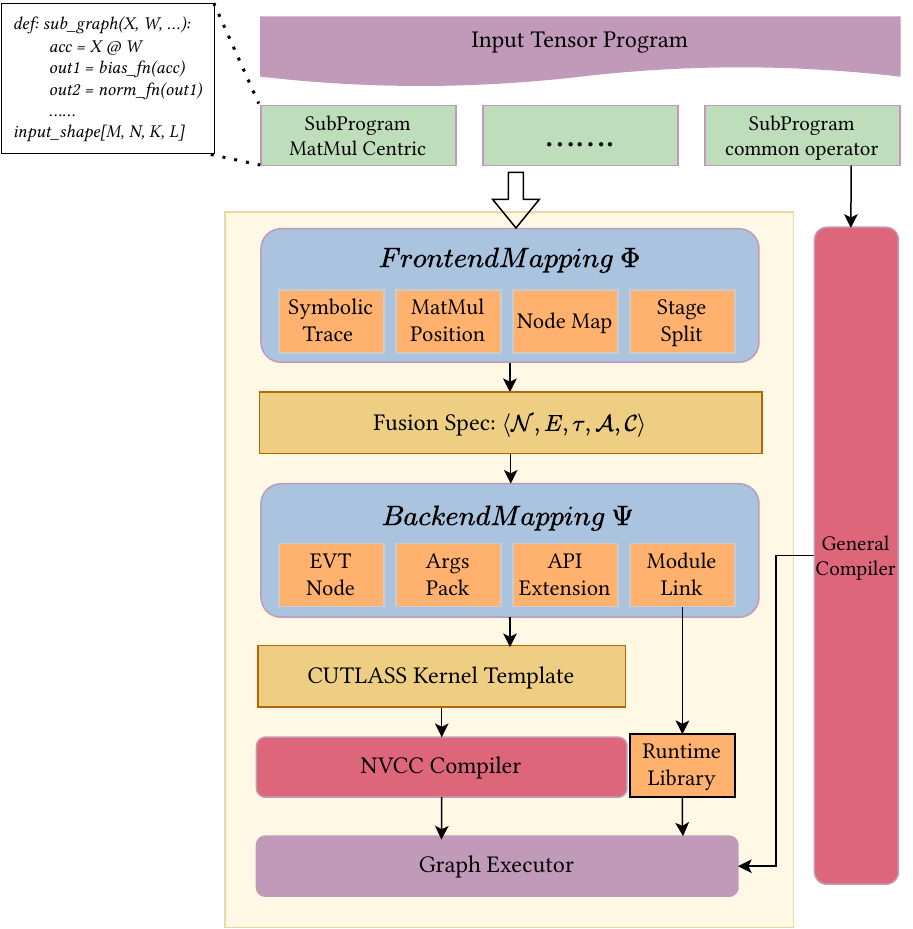}
  \caption{ComFuse Compilation Stack.}
  \label{fig:5_1}
\end{figure}

\subsection{Code Translate} 
The \textit{ComFuse} system translates a tensor subprogram across three distinct states, as illustrated in Figure~\ref{fig:5_1}. 
Initially, the frontend captures the Python-defined subgraph as a symbolic graph and localizes the compute-intensive and memory-intensive portions, translating the program into the the Fusion Spec IR, which unifies the memory-intensive subgraphs into a topologically ordered list of nodes. 
Finally, the backend renders this IR into the third state—the target template code—and invokes the compilation pipeline. 
We formalize the translation process as a two-step homomorphic mapping:
\begin{equation}
  G_{\text{epi}} \;\xrightarrow{\Phi}\; \mathcal{S} \;\xrightarrow{\Psi}\; \mathcal{T},
\end{equation}
where $G_{\text{epi}}$ is the Epilogue subgraph extracted from the original computation graph, $\mathcal{S}$ represents the intermediate Fusion Spec IR, and $\mathcal{T}$ denotes the generated template code of the target kernel.

\subsection{Frontend Mapping} 
Given the captured graph $G_{\text{epi}}$, the mapping $\Phi$ automatically initiates a top-down traversal of the subgraph starting from the MatMul node $v_{\text{mm}}$ as the root, systematically transforming the raw computational nodes and their dependencies into the formalized Fusion Spec IR.
It applies a set of structured patterns to each node: arithmetic binary operations are mapped to “broadcast + computation” pairs, constants are materialized as scalar broadcast nodes, one-dimensional placeholder tensors are recognized as row-vector broadcasts and automatically registered as runtime parameters, and unary functions (including activations, rsqrt, square, etc.) are directly mapped to corresponding computation nodes. We formalize Fusion Spec IR as a five-tuple:
\begin{equation}
  \mathcal{S} = \langle \mathcal{N}, E, \tau, \mathcal{A}, \mathcal{C} \rangle,
\end{equation}
where $\mathcal{N}$ denotes the topologically ordered sequence of nodes (encompassing semantic categories such as accumulator entry, broadcast, element-wise computation, and reduction), $E$ represents the data dependency edges between nodes, $\tau$ indicates the terminal node, $\mathcal{A}$ is the table of external runtime parameters, and $\mathcal{C}$ optionally captures nested sub-reductions.
We deliberately restrict the domain of $  \Phi  $: operators falling outside this pattern set are fallback to the standard compilation pathway.
This closure property ensures that the translation process does not silently drop semantics — any subgraph accepted by $  \Phi  $ has its semantics fully preserved in $  \mathcal{S}  $.

\subsection{Reduction Subgraph Splitting} 
The Stage-Stream Execution Model partitions the memory-intensive subgraph into a chain of sub-DAGs, which are embedded as PostOp template parameters within their preceding Reduction nodes.
We address this through a recursive construction: upon encountering a Reduction node $  v_r  $, $  \Phi  $ triggers a subgraph split on $  G_{\text{epi}}  $ with $  v_r  $ as the new root, encapsulates all nodes following the Reduction into a nested $  \mathcal{S}  $, and introduces necessary placeholder entry points to carry the reduction statistics and the passthrough original tensor, respectively.
By tracing back the original input data of the Reduction node, semantics such as “reduce over inputs but normalize using original values” (as in LayerNorm) are naturally expressed.
Consequently, common patterns including LayerNorm, RMSNorm, and Softmax can be uniformly converted into recursively nested $  \mathcal{S}  $ of bounded depth without requiring any pattern-specific recipes.

\subsection{Backend Mapping}
The mapping $\Psi$ performs a structural traversal over the Fusion Spec IR $\mathcal{S}$, deterministically expanding each node category into its corresponding EVT template instantiation. 
This expansion operates via structural induction: the data dependency edges of each node are mapped to a type-level topological description within the CUTLASS type system, while runtime parameters are assembled into a recursively constructed initialization sequence. 
Nested IR structures recursively trigger this same expansion process, embedding themselves into the outer visitor scope. 

Because this template rendering is entirely data-driven, the mapping $\Psi$ is highly extensible and open-ended at the node category level. 
Introducing a new semantic operator requires only extending the IR vocabulary and its corresponding expansion rules, enabling seamless support for new operators without modifying the frontend parser or the scheduling engine.

\section{Evaluation}
\label{sec:evaluation}
We conduct microbenchmarks on representative subgraph structures commonly found in deep learning workloads to evaluate the performance of fusion kernels generated by the \textit{ComFuse} system, and compare them against state-of-the-art general-purpose compiler systems. 

Our evaluation focuses on two aspects. 
First, we investigate the effectiveness of the Stage-Stream Execution Model under structures consisting of a MatMul followed by memory-intensive subgraphs. 
Second, we further evaluate the capability of \textit{ComFuse} to support B2BGEMM fusion for more complex subgraph patterns. 
Finally, based on the empirical results, we analyze the advantageous scenarios and inherent limitations of the \textit{ComFuse} system.

\begin{figure*}[t]
  \centering
  \includegraphics[width=\textwidth]{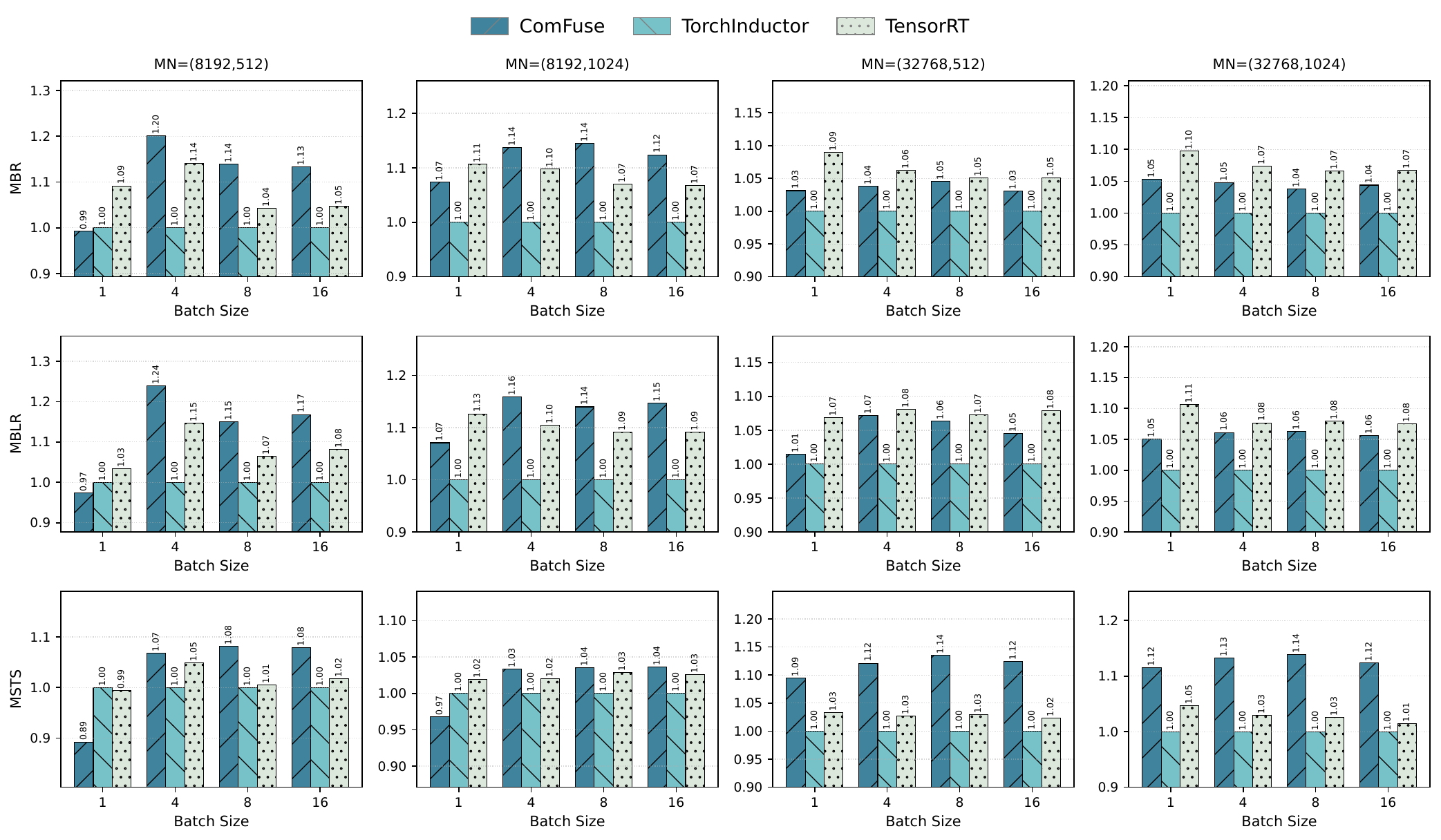}
  \caption{Overall performance comparison for Stage-Stream Execution Model.}
  \label{fig:evaluation-stream-reduction-overall}
\end{figure*}

\paragraph{Setup} 
The experimental environment consists of Python 3.10.19, PyTorch 2.5.1, and CUDA 12.4. To eliminate the fixed host-side kernel launch overhead, we employ CUDA Graphs to capture and replay kernel executions, and measure the corresponding GPU execution time. 
For each kernel configuration, we perform 50 warm-up iterations followed by 500 measured iterations, and report the averaged runtime results.

\paragraph{Baselines} 
We compare \textit{ComFuse} against two representative compiler-based optimization frameworks.

\textbullet\ \textbf{TorchInductor:} The default PyTorch 2.0 compiler backend, which dispatches computations to vendor libraries or generates customized fused kernels via Triton to achieve near-expert performance.

\textbullet\ \textbf{TensorRT:} NVIDIA's high-performance inference engine, which serves as a strong industrial baseline by leveraging graph-level optimizations, operator fusion, and kernel auto-tuning for MatMul-centric subgraphs.

For TorchInductor, we enable static-shape compilation and \textit{max-autotune} in \texttt{torch.compile} to obtain a strong subgraph-level baseline. For TensorRT, we employ NVIDIA TensorRT 10.3.0 as the inference engine. 
All benchmark workloads are first exported to ONNX computational graphs, which are then imported, optimized, and compiled by TensorRT. 
Full static-shape optimizations are enabled throughout the compilation pipeline to ensure that TensorRT achieves its peak achievable performance, providing a fair and representative baseline for comparison.

\subsection{Performance for Stage-Stream Execution Model}
\label{sec:stream-reduction-fusion-performance}
\subsubsection{Workloads} We select three representative patterns to evaluate the performance of the Stage-Stream Execution Model in \textit{ComFuse} for fusing MatMul followed by complex memory-intensive subgraphs.
First, \textit{MatMul-BiasAdd-RMSNorm} (MBR) represents a relatively simple reduction-based, memory-intensive epilogue pattern. 
Second, \textit{MatMul-BiasAdd-LayerNorm-ReLU} (MBLR) combines linear transformation, normalization, and nonlinear activation, introducing a more complex epilogue. 
Third, \textit{MatMul-Scale-TanhSoftcap-Softmax} (MSTS) corresponds to attention-score normalization, where the epilogue involves both nonlinear transformation and a computationally intensive Softmax reduction.

\begin{figure*}[t]
  \centering
  \includegraphics[width=\textwidth]{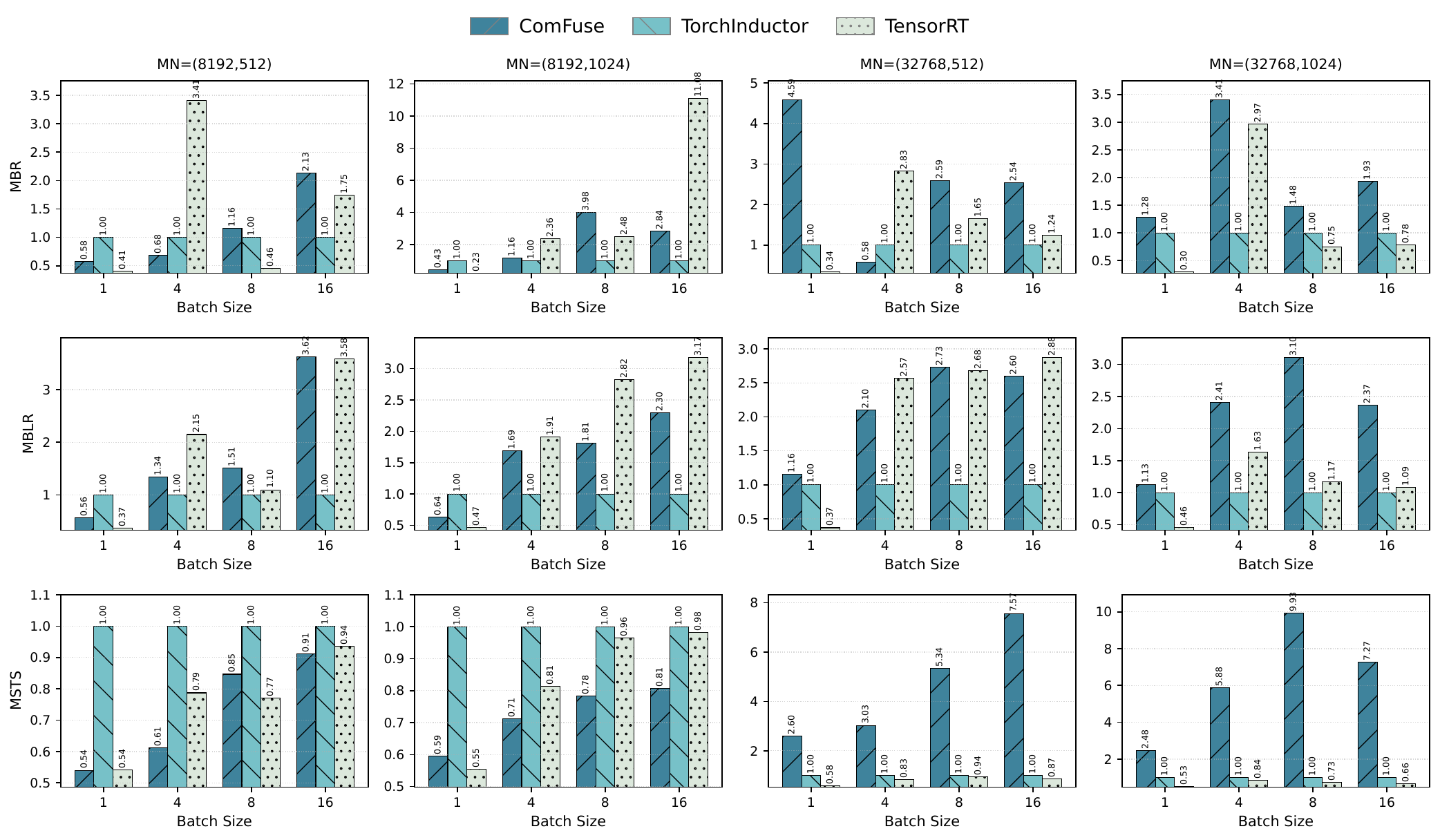}
  \caption{Residual-time performance comparison for Stage-Stream Execution Model, as Residual-time is defined as $\mathrm{Time}(\mathrm{Total}) - \mathrm{Time}(\mathrm{MatMul})$.}
  \label{fig:evaluation-stream-reduction-residual}
\end{figure*}

\subsubsection{Overall Results}
Figure~\ref{fig:evaluation-stream-reduction-overall} presents the end-to-end performance of \textit{ComFuse} relative to the baseline systems, where the execution time of TorchInductor is normalized to 1.0.
Overall, \textit{ComFuse} consistently outperforms TorchInductor across all evaluated workloads, achieving up to 1.24× speedup.
Under the three patterns, \textit{ComFuse} delivers average speedups of 1.08×, 1.09×, and 1.07×, respectively.
When compared against TensorRT, which represents a highly optimized inference engine, \textit{ComFuse} achieves comparable or better performance in most evaluated scenarios.
This demonstrates that the proposed execution model and fusion strategy remain competitive even against mature inference optimization frameworks.

A closer examination reveals that, for workloads with relatively small computational scales (e.g., small problem sizes or batch sizes), \textit{ComFuse} may slightly underperform baseline on certain subgraphs. 
This behavior can be attributed to the limited number of tiles assigned to each CTA, preventing the pipeline from reaching a steady-state execution regime. 
In such cases, the scheduling and coordination overhead introduced by the fused kernel constitutes a relatively larger fraction of the overall execution time, thereby reducing the benefits of joint fusion.
As the computational workload increases, each CTA processes a larger number of tiles, enabling the pipeline to effectively hide the execution latency of the memory-intensive subgraphs. 
Consequently, \textit{ComFuse} achieves superior pipeline efficiency and demonstrates a widening performance advantage over the baselines.

\begin{table}[!t]
  \caption{Computational complexity of representative Epilogue patterns.}
  \label{tab:epilogue-complexity}
  \centering
  \small

  \begin{tabular}{@{}p{0.16\columnwidth}p{0.24\columnwidth}p{0.48\columnwidth}@{}}
    \toprule
    \textbf{Pattern} & \textbf{CUDA ALU} & \textbf{CUDA SFU} \\
    \midrule
    MBR  & $M(4N+1)$ & $M \cdot \mathrm{rsqrt}$ \\
    MBLR & $M(8N+1)$ & $M \cdot \mathrm{rsqrt} + MN \cdot \mathrm{comp}$ \\
    MSTS & $M(6N-1)$ & $MN \cdot (\tanh + \exp) + M(N-1) \cdot \mathrm{comp}$ \\
    \bottomrule
  \end{tabular}
\end{table}

Table~\ref{tab:epilogue-complexity} summarizes the computational complexity of the three patterns. 
Combined with the results in Figure~\ref{fig:evaluation-stream-reduction-overall}, we observe that the speedup achieved by \textit{ComFuse} generally increases with the computational intensity of the memory-intensive subgraph. 
In particular, the MSTS pattern, which exhibits the highest computational complexity, achieves the largest performance improvement. 
This result demonstrates that when the pipeline operates at peak efficiency, the fused kernels generated by \textit{ComFuse} exhibit a more pronounced advantage for complex, memory-intensive subgraphs. 
This superiority stems from the pipeline's capability to successfully overlap the time-consuming epilogue computations under the shadow of MatMul execution.

Since the MatMul operation dominates the overall execution time, end-to-end performance alone cannot fully isolate the benefits of our joint optimization. 
Therefore, we define the residual time as $\mathrm{Time}(\mathrm{Total}) - \mathrm{Time}(\mathrm{MatMul})$ to evaluate the memory-intensive subgraph under both standalone and fused execution (Figure~\ref{fig:evaluation-stream-reduction-residual}). 
For the MBR and MBLR patterns, the speedup of \textit{ComFuse} scales with the batch size, benefiting from pipeline steady-state. 
Conversely, for the MSTS pattern, \textit{ComFuse} underperforms the baselines in several small-scale test cases. 
This is primarily due to the workload imbalance between the MatMul and epilogue stages, which limits the extent to which epilogue computations can be hidden behind MatMul execution. 
Nevertheless, as the workloads of the two stages become more balanced, \textit{ComFuse} delivers substantial performance improvements, achieving up to a 9.93$\times$ speedup.

\subsection{Performance for B2BGEMM Fusion}
\subsubsection{Workloads}
We evaluate \textit{ComFuse} on three representative B2BGEMM workloads, covering both attention-based and MLP-based patterns.
The Self-Attention~\cite{vaswani2017attention} workload represents the dense attention computation commonly used in Transformer models, while the Target-Attention~\cite{zhou2018deepinterestnetwork} workload captures asymmetric attention with distinct target and source sequence lengths.
The DLRM Bottom MLP\cite{naumov2019deep} workload represents consecutive matrix multiplications in recommendation models, where feature dimensions are progressively reduced across layers.
The detailed experimental dimensions and configurations are summarized in Table~\ref{tab:workloads}. 
Together, these workloads allow us to assess the performance of \textit{ComFuse} under diverse matrix shapes and operator compositions.

\begin{table}[t]
\caption{Evaluated B2BGEMM workloads.}
\label{tab:workloads}
\begin{tabular}{ll}
\toprule
\textbf{Pattern} & \textbf{Parameters} \\
\midrule
Self-Attention & $\text{SeqLen}=512,\ \text{HeadDim}=128$ \\
Target-Attention & $\text{TargetLen}=4096,\ \text{SourceLen}=256$ \\
DLRM Bottom MLP & $\text{FeatureDim}: 1024 \rightarrow 512 \rightarrow 128$ \\
\bottomrule
\end{tabular}
\end{table}

\begin{figure}[t]
  \centering
  \includegraphics[width=\columnwidth]{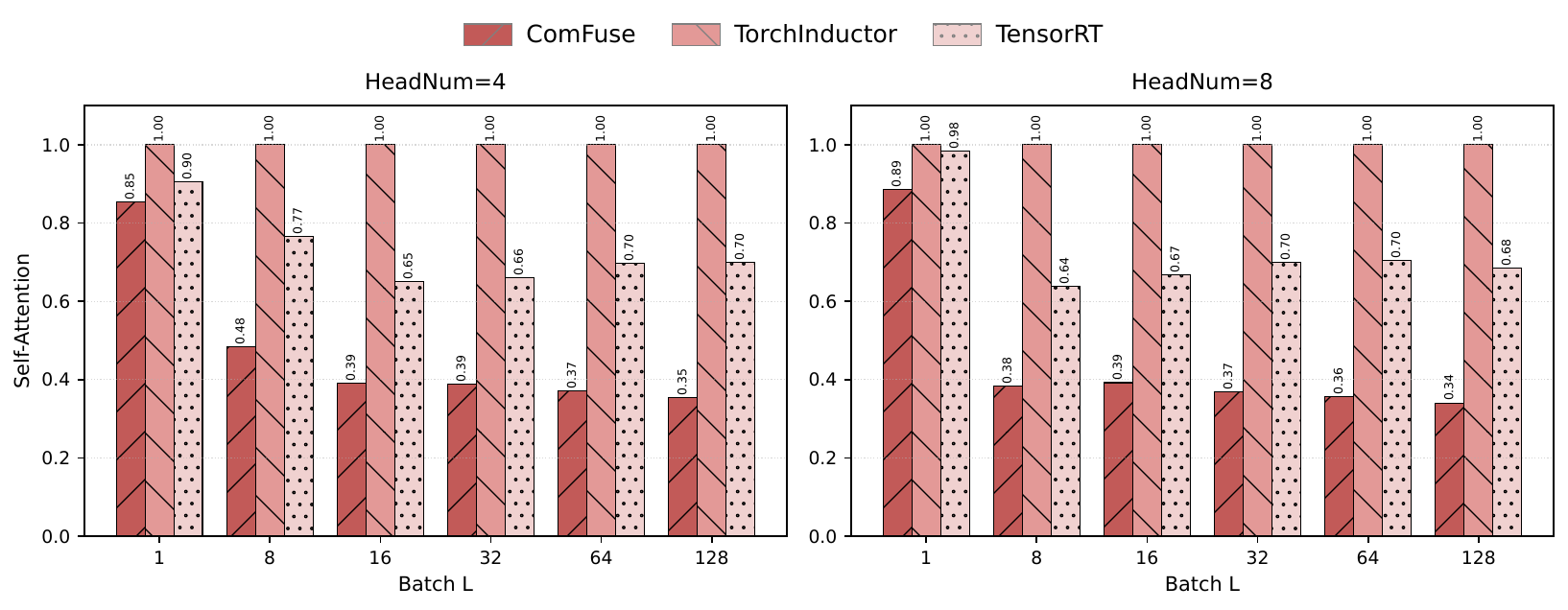}
  \caption{Performance for Self-Attention.}
  \label{fig:self-attn}
\end{figure}
\begin{figure}[t]
  \centering
  \includegraphics[width=\columnwidth]{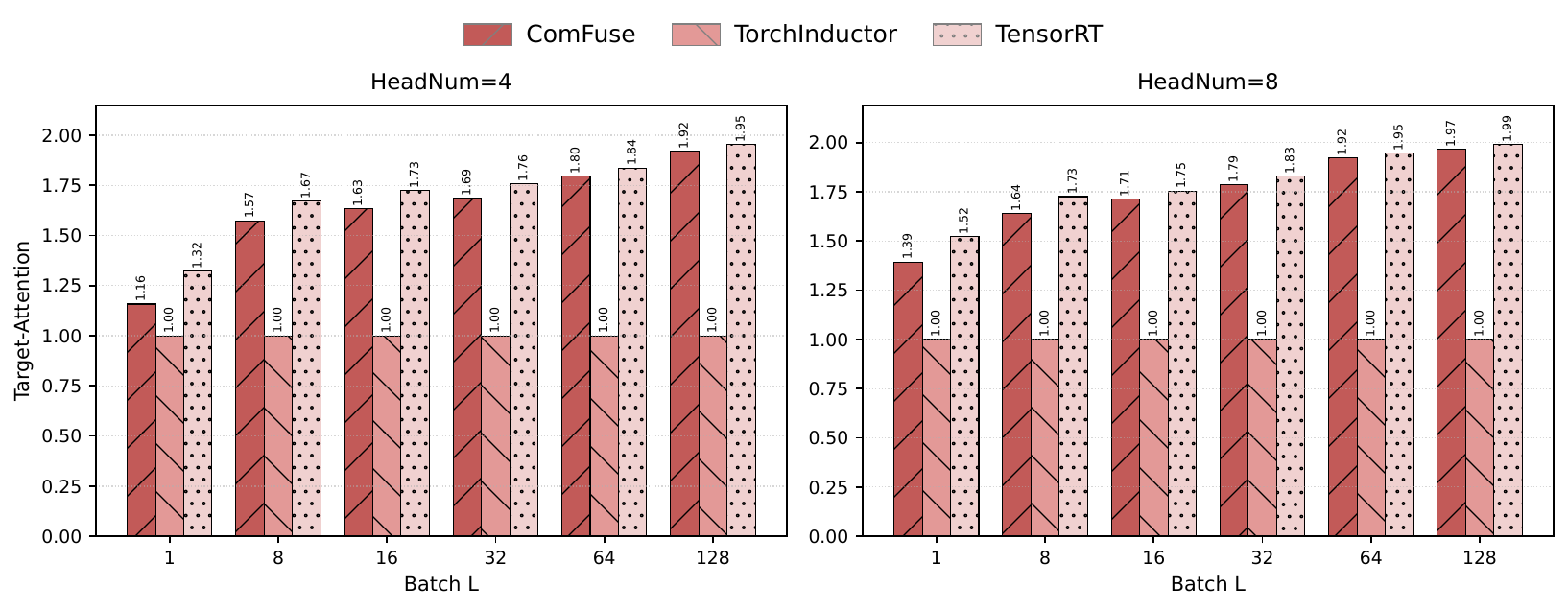}
  \caption{Performance for Target-Attention.}
  \label{fig:target-attn}
\end{figure}
\begin{figure}[t]
  \centering
  \includegraphics[width=\columnwidth]{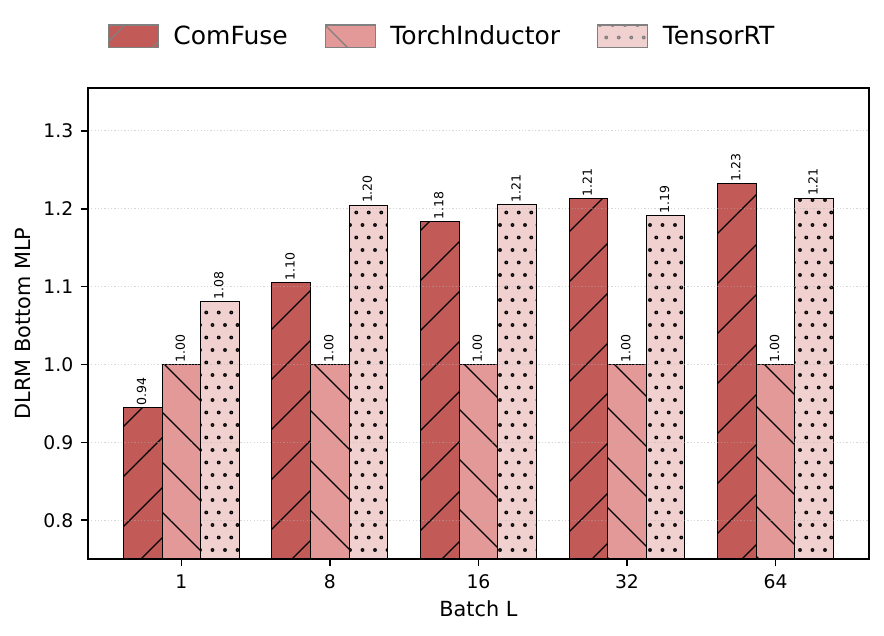}
  \caption{Performance for DLRM Bottom MLP.}
  \label{fig:MLPTower}
\end{figure}

\subsubsection{Overall Results}
Figure~\ref{fig:self-attn}, \ref{fig:target-attn}, and~\ref{fig:MLPTower} presents the performance comparison between \textit{ComFuse} and the baseline across the evaluated workloads.
In the Target-Attention and DLRM Bottom MLP scenarios, compared with the TorchInductor baseline, \textit{ComFuse} achieves up to a 1.97× speedup in the Target-Attention and up to a 1.23× speedup in the DLRM Bottom MLP, while delivering performance close to that of TensorRT-optimized execution.It is worth noting that the optimization capability of TensorRT largely relies on graph-level pattern matching. For regular structures and standard operator compositions, TensorRT can trigger highly optimized kernels through predefined patterns. However, its applicability is often limited when the model contains complex structures, non-standard operator compositions, or structural variants. 
In contrast, \textit{ComFuse} does not depend on fixed graph-level patterns; instead, it performs more general fusion and scheduling optimizations for B2BGEMM computation patterns. Therefore, \textit{ComFuse} offers stronger applicability and extensibility for complex or variant model structures.

In the Self-Attention scenario, we observe that TorchInductor successfully matches its built-in FlashAttention\cite{dao2023flashattention2} operator library. As a result, the Torch baseline achieves the best performance in this case.
In contrast, \textit{ComFuse} exhibits a substantial performance gap compared to the baseline in this scenario. 
We attribute this behavior primarily to the imbalance in computational workload between the MatMul stage and the Softmax stage.
Specifically, in Self-Attention, the time complexity of the MatMul component is approximately $O(N^2 d)$, where $N$ denotes the sequence length and $d$ denotes the attention head dimension. The time complexity of the Softmax stage is approximately $O(N^2)$. 
Although Softmax has lower asymptotic complexity than MatMul, its \texttt{exp}, reduction, and normalization operations are less amenable to Tensor Core acceleration and typically execute on ALU and SFU. As a result, the Softmax stage can dominate a larger fraction of the runtime, leading to an imbalanced pipeline.

This observation also provides an important implication for the design of \textit{ComFuse}: directly fusing compute-intensive operators with memory-intensive operators does not necessarily lead to optimal performance. An effective fusion strategy must further account for the computational characteristics of different stages, the heterogeneity of hardware execution units, and the balance of pipeline workloads.

\section{Conclusion and Limitation}
\label{sec:conclusion}
Operator fusion is a key optimization in deep learning compilation; however, existing compilation stacks fail to effectively resolve the data boundaries between compute-intensive and memory-intensive operators. 
We observe that fusing these two distinct classes of operators yields substantial potential benefits in both memory bandwidth and computational efficiency. 
We propose \textit{ComFuse}, a compilation system that leverages modern GPU architectural features to automatically captures computational semantics within tensor subprograms, and fuses complex memory-intensive subgraphs and compute-intensive operators into high-performance kernels. 
Specifically, \textit{ComFuse} introduces a Stage-Stream Execution Model to break the execution boundaries imposed by reductions, enabling pipelined cooperation across computation stages and reducing intermediate data movement. 
It further supports complex B2BGEMM fusion structures, thereby extending the applicability of conventional fusion paradigms. 
Experimental results demonstrate that \textit{ComFuse} achieves performance comparable to or better than state-of-the-art compilation approaches on representative workloads, while providing stronger applicability to complex subgraphs and structural variants.

Although \textit{ComFuse} has achieved substantial performance gains in various scenarios, it still exhibits certain limitations, particularly in lacking more sophisticated optimization strategies tailored for compute-intensive operators. 
Under specific workloads, a performance gap remains between \textit{ComFuse} and state-of-the-art baselines. 
Nevertheless, analyzing these cases has deepened our understanding of the underlying mechanisms governing compute-memory joint operator fusion. 
Moving forward, we plan to deepen our analysis of operator computational characteristics and hardware architectural features, enabling the system to adaptively select the most suitable optimization strategies across diverse workloads.


\bibliographystyle{IEEEtranS}
\bibliography{refs}

\end{document}